\documentclass[%
 aip,
 amsmath,amssymb,
 reprint,%
]{revtex4-1}

\usepackage{graphicx}
\usepackage{dcolumn}
\usepackage{float}
\usepackage{placeins}
\usepackage{bm}
\usepackage[utf8]{inputenc}
\usepackage[T1]{fontenc}
\usepackage{mathptmx}
\usepackage{color}

\begin{document}

\preprint{AIP/123-QED}

\title[r$^2$SCAN-D4: Dispersion corrected meta-generalized gradient approximation for general chemical applications]{r$^2$SCAN-D4: Dispersion corrected meta-generalized gradient approximation for general chemical applications}

\author{Sebastian Ehlert}
\affiliation{Mulliken Center for Theoretical Chemistry, University of Bonn, Beringstr. 4, 53115 Bonn, Germany}
\author{Uwe Huniar}
\affiliation{Biovia, Dassault Syst\`{e}mes Deutschland GmbH,
Imbacher Weg 46, 51379 Leverkusen, Germany}
\author{Jinliang Ning}
\affiliation{Department of Physics and Engineering Physics, Tulane University, New Orleans, Louisiana 70118, United States}
\author{James W. Furness}
\affiliation{Department of Physics and Engineering Physics, Tulane University, New Orleans, Louisiana 70118, United States}
\author{Jianwei Sun}
\affiliation{Department of Physics and Engineering Physics, Tulane University, New Orleans, Louisiana 70118, United States}
\author{Aaron D. Kaplan}
\affiliation{Department of Physics, Temple University, Philadelphia, Pennsylvania 19122, United States}
\author{John P. Perdew}
\affiliation{Department of Physics, Temple University, Philadelphia, Pennsylvania 19122, United States}
\affiliation{Department of Chemistry, Temple University, Philadelphia, Pennsylvania 19122, United States}
\author{Jan Gerit Brandenburg}
\homepage{gerit-brandenburg.de}
\email{j.g.brandenburg@gmx.de}
\affiliation{
Enterprise Data Office,
Merck KGaA, Frankfurter Str. 250, 64293 Darmstadt, Germany}

\date{\today}%

\begin{abstract}
We combine a regularized variant of the strongly constrained and appropriately normed semilocal density functional [J. Sun, A. Ruzsinszky, and J. P. Perdew, \emph{Phys. Rev. Lett.} {\bf 115}, 036402 (2015)] with the latest generation semi-classical London dispersion correction. The resulting density functional approximation r$^2$SCAN-D4 has the speed of generalized gradient approximations while approaching the accuracy of hybrid functionals for general chemical applications.
We demonstrate its numerical robustness in real-life settings and benchmark molecular geometries, general main group and organo-metallic thermochemistry, as well as non-covalent interactions in supramolecular complexes and molecular crystals. Main group and transition metal bond lengths have errors of just 0.8\%, which is competitive with hybrid functionals for main group molecules and outperforms them for transition metal complexes. The weighted mean absolute deviation (WTMAD2) on the large GMTKN55 database of chemical properties is exceptionally small at 7.5 kcal/mol. This also holds for metal organic reactions with an MAD of 3.3 kcal/mol. The versatile applicability to organic and metal-organic systems transfers to condensed systems, where lattice energies of molecular crystals are within chemical accuracy (errors $<$1\,kcal/mol).
\end{abstract}

\maketitle

\section{\label{sec:intro}Introduction}

The quantum mechanical description of physical and chemical materials at electronic resolution is an increasingly important task for \textit{in silico} simulations.
Here, density functional theory (DFT) has emerged in the past decades as one of the most versatile methodological frameworks.\cite{dftbook,kohn98}
This leading position in both materials and chemical applications is largely due to the excellent accuracy over computational cost ratio, as well as the broad applicability across system classes of today's density functional approximations (DFAs).\cite{Burke2012:jcp_rev, Becke-JCP, dft-materials-rev}

The Jacob's ladder hierarchy \cite{JacobsLadder} is commonly used to classify DFAs.
In this hierarchy, DFAs are systematically improved by ascending rungs of different approximations: the local density approximation (LDA), generalized gradient approximations (GGAs), meta-GGAs, hybrid functionals (including a fraction of nonlocal exact exchange), and double-hybrid functionals (including nonlocal correlation).
In terms of efficiency, meta-GGAs are in a favorable spot, as they have the same cubic scaling with system size as LDA.
Yet, many of the meta-GGAs proposed so far cannot truly leverage the full potential of their rung.
Some shortcomings of existing functionals are increased sensitivity to the numeric integration grid, as observed in the SCAN functional\cite{scan} or several Minnesota type functionals,\cite{m06l,m11l,minnesota_bsse} purely empirical parameters, as present in the B97M functional\cite{mardirossian2015}, and sensitivity to the kinetic energy density.\cite{minnesota_raregas,mgga_oscillations}
Recent developments of semi-local DFAs combine exact constraints with various degrees of parametrization to improve descriptions of short- to medium- range electron correlation~\cite{scan,wb97xv,revm06l}.

The strongly constrained and appropriately normed (SCAN) functional\cite{scan} is constructed to rigorously satisfy all known exact constraints for meta-GGAs.
While the functional itself has shown excellent performance in previous studies, the severe numerical instabilities inherent to the functional impeded its adoption for many computational studies.
With the recently proposed regularized SCAN (rSCAN) \cite{bartok2019} and the subsequent restoration of exact constraints in r\textsuperscript2SCAN \cite{furness2020}, the main drawback of the SCAN functional seems to be resolved.

Nevertheless, semilocal functionals cannot include long-range correlation effects like London dispersion interactions.\cite{disp_chemrev}
To truly judge its applicability, we extensively tested r\textsuperscript2SCAN combined with the state-of-the-art D4 dispersion correction\cite{caldeweyher2019}, which shows unprecedented performance for a range of diverse chemical and physical properties.
To investigate the development of the SCAN-type functionals we include both SCAN-D4 and rSCAN-D4 in the comparison to r\textsuperscript2SCAN-D4, and can attribute  improvements in non-covalent interactions mainly to the regularization and improvements for thermochemistry and barrier heights to the restoration of the exact constraints.

We give a concise methodological overview (Section~\ref{sec:methods}) on r\textsuperscript2SCAN and D4 before testing the full method against established DFAs over a wide range of benchmarks (Section ~\ref{sec:results}), with particular focus on molecular geometries, thermochemistry, kinetics, and non-covalent interactions in small and large complexes.


\section{\label{sec:methods}Methods}

The rSCAN\cite{bartok2019} functional regularizes the severe numerical instability or inefficiency of the otherwise successful SCAN\cite{scan} functional at the expense of breaking exact constraints SCAN was constructed to obey. This problem arises in many codes that employ localized basis sets, and is less problematic in many codes that employ plane-wave basis sets. While numerical challenges are indeed resolved, a rigorous adherence to exact constraints is core to the design of the SCAN functional and likely important for transferable accuracy across  domains of applicability.\cite{mejia-rodriguez2020a} This seems to be reflected in rSCAN's relatively poor performance for molecular atomization energies compared to other tests.\cite{Mejia-Rodriguez2019, Bartok2019a} The r\textsuperscript2SCAN functional\cite{furness2020} combines the good accuracy of SCAN with the numerical efficiency of rSCAN by directly restoring exact constraint satisfaction to the rSCAN regularizations. 


The SCAN functional is constructed as an interpolation between single orbital and slowly-varying energy densities designed to maximize exact constraint satisfaction.\cite{scan} The interpolation is controlled by an iso-orbital indicator
\begin{equation}
    \alpha = \frac{\tau - \tau_\mathrm{W}}{\tau_\mathrm{U}},
\end{equation}
where $\tau_\mathrm{W} = |\nabla \rho|^2/(8\rho)$ and $\tau_\mathrm{U} = 3(3\pi^2)^{2/3}\rho^{5/3}/10$ are the von-Weizs\"acker and uniform electron gas kinetic energy densities respectively.\cite{Sun2013a} In subsequent studies, $\alpha$ has been shown to contribute to numerical instability.\cite{scand3,Furness2019} To remove these effects, a regularized $\alpha^\prime$ was used in rSCAN that removes single orbital divergences at the expense of breaking exact coordinate scaling conditions\cite{Levy1985a, Gorling1993a, Pollack2000} and the uniform density limit. These conditions are restored in r\textsuperscript2SCAN by adopting a different regularization:
\begin{equation}
    \bar{\alpha} = \frac{\tau - \tau_\mathrm{W}}{\tau_\mathrm{U} + \eta\tau_\mathrm{W}},
\end{equation}
where $\eta = 10^{-3}$ is a regularization parameter.

The second regularization made in the rSCAN functional is to substitute the twisted piece-wise exponential interpolation of the original SCAN with a smooth polynomial function. This removes problematic oscillations in the exchange-correlation potential, but introduces spurious terms in the slowly-varying density gradient expansion that deviate from the exact expansion\cite{Svendsen1996, Perdew1992a} recovered by SCAN. A corrected gradient expansion term is used in r\textsuperscript2SCAN that cancels these spurious terms so the functional recovers the slowly-varying density gradient expansion to second order.
A recent modification of SCAN for improved band gap accuracy from Aschebrock and K\"ummel named ``TASK'' \cite{Aschebrock2019} is able to enforce the fourth-order gradient expansion for the exchange energy without apparent numerical problems\cite{Hofmann2020}, although TASK uses an LSDA for correlation. The importance of the fourth-order exchange terms is not established however, and we are thus satisfied using one less exact constraint compared to SCAN.

\subsection{\label{subsec:numeric}Numerical stability}

\begin{figure}[htb]
\includegraphics[width=0.45\textwidth]{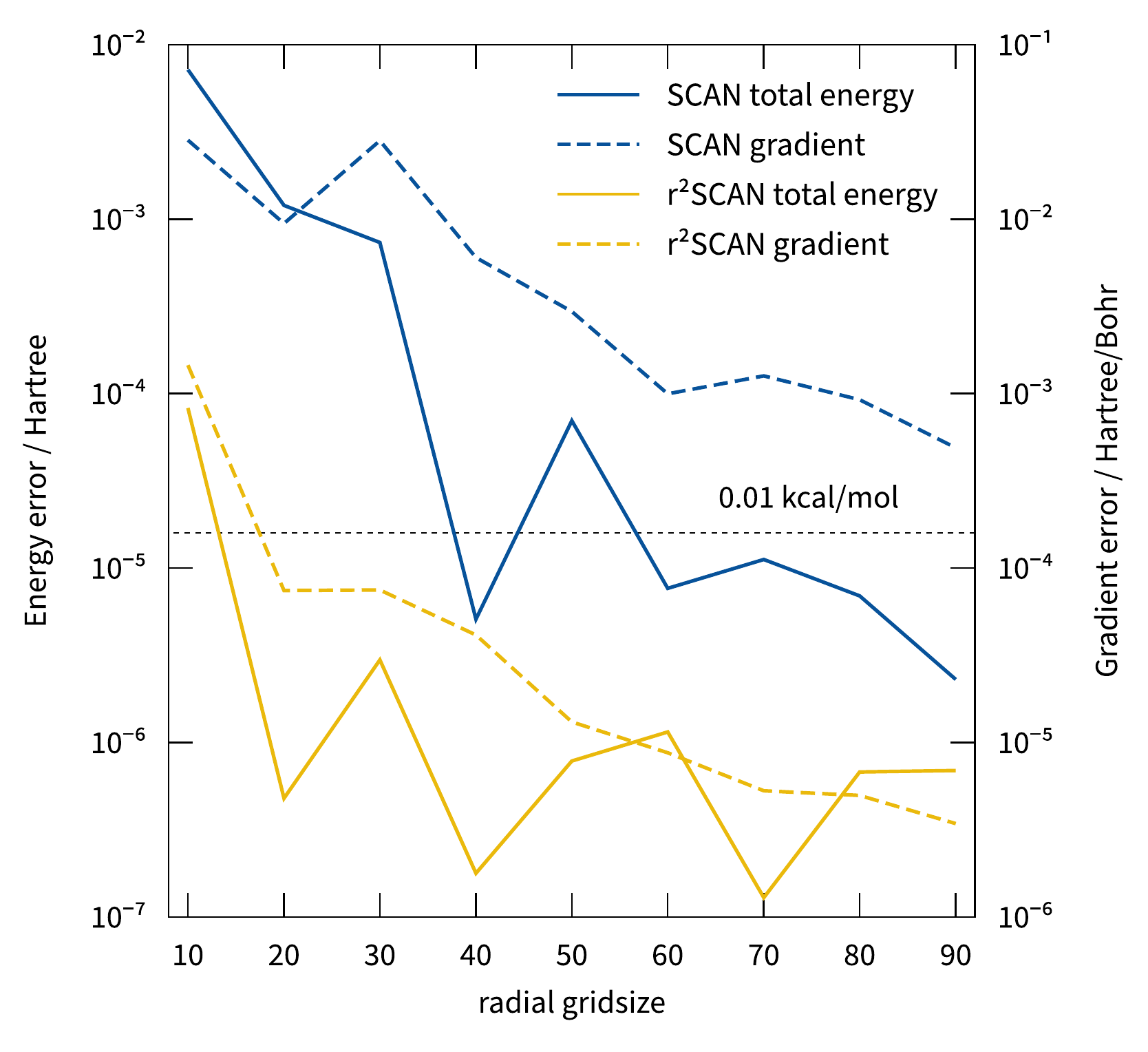}
\vspace{-0.4cm}
\caption{\footnotesize Errors for FeCp\textsubscript2 with SCAN/def2-QZVP and r\textsuperscript2SCAN/def2-QZVP using different radial gridsizes.
For both methods, grid 4 and SCF convergence criteria of $10^{-7}$\,Hartree were used and the radial gridsize was varied. The reference has a radial gridsize of 100.
The gradient error is the sum of the absolute errors of all gradient components. For further explanation, see the end of Section \ref{subsec:numeric}.
}
\label{fig:numeric}
\end{figure}


Numerical instabilities are revealed by SCAN's sensitivity to the choice of numerical integration grid, often requiring dense, computationally costly grids.\cite{scand3,bartok2019,mejia-rodriguez2020b}
This issue has been addressed with the rSCAN and r\textsuperscript2SCAN functionals. Fig.~\ref{fig:numeric} shows that the regularization indeed leads to two orders of magnitude error reduction when comparing r\textsuperscript2SCAN with SCAN. This holds for both total energy and nuclear gradients for all chosen numerical settings.
In practice, this allows for more computationally favorable settings.
To give a rough estimate of the computational cost of r\textsuperscript2SCAN compared to SCAN, we consider system 10 of the S30L\cite{s30l} with 158 atoms and 8250 atomic orbitals in a def2-QZVP basis set.
A SCAN calculation using Turbomole's grid 4 and radsize 50 would take approximately 10~hours, while an r\textsuperscript2SCAN calculation with Turbomole grid m4 and radsize 6 takes only three and a half hours for the same numerical accuracy, resulting in a computational saving of a factor of three to five.\footnote{Running on Intel(R) Xeon(R) CPU E3-1270 v5 @ 3.60GHz using four cores.}
We recommend using r\textsuperscript2SCAN with 6 radial points and potentially increasing it to 10 for problematic geometry optimizations.\footnote{The 6 radial points correspond to the default settings of Turbomole's \emph{grid m4}.}

\subsection{\label{subsec:damping}Training of damping functions}
As London dispersion interactions arise from  nonlocal electron correlations, they cannot be captured by any meta-GGA. In the past years, a range of schemes have been developed to capture these interactions in the DFT framework.\cite{vdw_perspective,disp_chemrev, chemrev_tkatchenko, vdwdf_review,vydrov2010,sabatini2013} Here, we combine r\textsuperscript2SCAN with the semi-classical D4 dispersion correction.\cite{caldeweyher2019}
Its energy contribution is calculated by
\begin{equation}
\begin{split}
  E_{\text{disp}}^{\text{D4}} = 
  -\frac12\sum_{n=6,8}\sum_{\text A,\text B}^{\text{atoms}} s_n\frac{C_n^\text{AB}}{R_\text{AB}^n}\cdot f_n^\text{BJ}(R_\text{AB}) \\
  - \frac16\sum_{\text A,\text B,\text C}^{\text{atoms}} s_9\frac{C_9^\text{ABC}}{R_\text{ABC}^9}\cdot f_9^\text{BJ}(R_\text{ABC},\theta_\text{ABC})
  , 
\end{split}
\end{equation}
where $R_\text{AB}$ is the atomic distance, $C^\text{AB}_n$ is the \textit nth-order dispersion coefficient, and $f^\text{BJ}_n(R_\text{AB})$ is the Becke--Johnson damping function\cite{bjdamp2,bjdamp3}.
$R_\text{ABC}$ and $C^\text{ABC}_9$ denote the geometrically averaged distance and dispersion coefficient, respectively, and $\theta_\text{ABC}$ is the angle dependent term of the triple-dipole contribution.\cite{atm_a,atm_b}
\begin{table}[bht]
\caption{D3(BJ) and D4 damping parameter for rSCAN and r$^2$SCAN functionals.}
\label{tab:damping}
\begin{ruledtabular}
\begin{tabular}{ll rrrr}
& model & s8 & a1& a2/Bohr & RMS\footnotemark[1]\\[0.1cm]
\cline{3-6}\\[-0.2cm]
rSCAN     & D3(BJ)-ATM  &  1.0886 & 0.4702 & 5.7341 & 0.31 \\
          & D4(EEQ)-ATM &  0.8773 & 0.4911 & 5.7586 & 0.30 \\[0.1cm]
r$^2$SCAN & D3(BJ)-ATM  &  0.7898 & 0.4948 & 5.7308 & 0.28 \\
          & D4(EEQ)-ATM &  0.6019 & 0.5156 & 5.7734 & 0.28
\end{tabular}
\end{ruledtabular}
\footnotetext[1]{Root-mean-square error in kcal/mol of dispersion corrected density functionals on the fit set S66x8\cite{s66}, S22x5,\cite{s22} and NCIBLIND.\cite{taylor2016}}
\end{table}
The $s_8$ parameter for the two-body dispersion and the $a_1$ and $a_2$ parameter entering the critical radius in the damping function are adjusted to match the local description of a specific DFA.
Damping parameters are fitted using a Levenberg--Marquardt least-squares minimization to reference interaction energies as described in Ref.~\citenum{caldeweyher2019}. Optimized parameters are given in Table~\ref{tab:damping}.

\subsection{\label{subsec:compdetail}Computational details}

All ground state molecular DFT calculations were performed with a development version of Turbomole.\cite{turbomole_wire,turbomole}
The resolution of identity (RI) approximation\cite{ridft,jauxbas} was applied in all calculations for the electronic Coulomb energy contributions.
For all functionals except SCAN, Turbomole's modified grids of type m4 were used. For all SCAN calculations, grid 4 with increased radial integration size of 50 was used instead.
Self-consistent field convergence criteria of $10^{-7}$\,Hartree were applied.
Ahlrichs' type quadruple-zeta basis sets, def2-QZVP,\cite{qzvp} were used throughout if not stated otherwise.

The periodic electronic structure calculations were conducted with {\sc Vasp\,6.1} ~\cite{vasp1,vasp2} with projector-augmented plane waves with an energy cutoff of 800 or 1000\,eV (hard PAWs~\cite{paw1,paw2}). Tight self-consistent field settings and large integration (and fine FFT) grids are used. The Brillouin zone sampling  has been increased to converge the interaction energy to 0.1\,kcal/mol. 
The non-periodic directions use a vacuum spacing of 12\,\AA.

\section{\label{sec:results}Results}

\subsection{\label{subsec:geo}Bond length and molecular geometries}

\begin{figure}[htb!]
\includegraphics[width=0.48\textwidth]{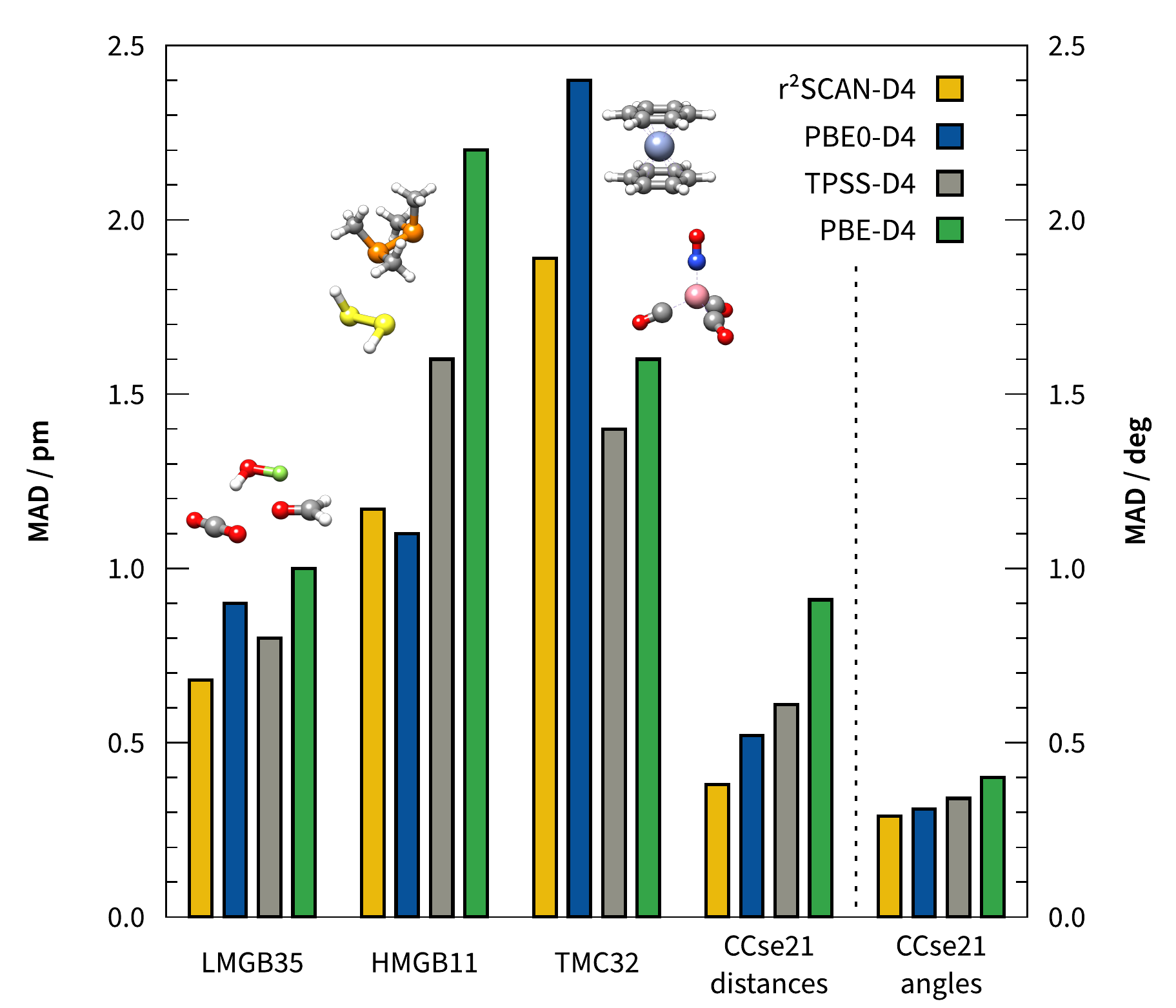}
\vspace{-0.4cm}
\caption{\footnotesize Errors in bond length from r$^2$SCAN-D4 and other DFAs separated into light main group bonds (LGMB35~\cite{pbeh3c}), heavy main group bonds (HMGB11~\cite{pbeh3c}), transition metal complexes (TMC32~\cite{tmc32}) and semi-experimental organic molecules (CCse21~\cite{ccse21}). PBE0-D4, TPSS-D4 and PBE-D4 results for the first three sets are taken from Ref.~\citenum{caldeweyher2019}.
}
\label{fig:bonds}
\end{figure}

To evaluate the description of covalent bond distances, we compare experimental and calculated ground-state equilibrium distances~$R_e$ (in pm) for 35 light main group bonds (LMGB35~\cite{pbeh3c}), 11 heavy main group bonds (HMGB11~\cite{pbeh3c}), and 50 bonds in 32 3d transition metal complexes (TMC32~\cite{tmc32}).
Additionally, we investigate the bond distances and angles for a set of simple organic molecules against accurate semi-experimental references.\cite{ccse21,ccse21_adamo}
Extended statistics and optimized geometries are made freely available.\footnote{Optimized r\textsuperscript2SCAN-D4/def2-QZVP geometries of the LMGB35, HMGB11,and TMC32 sets, statistical performance of the  LMGB35, HMGB11, TMC32, CCse21\cite{ccse21,ccse21_adamo}, GMTKN55, S30L, L7, C40x10 sets are provided at   https://github.com/awvwgk/r2scan-d4-paper}

We include r\textsuperscript2SCAN-D4, PBE0-D4\cite{pbe0}, TPSS-D4\cite{tpss}, and PBE-D4\cite{pbe} in the comparison shown in Fig.~\ref{fig:bonds}.
For organic molecules, we find exceptional performance for all functionals, with errors smaller than 1\,pm in the bond distances and half a degree in the bond angles.
While all methods reproduce the reference values closely, we observe the best agreement from r\textsuperscript2SCAN-D4  with a mean absolute deviation (MAD) of 0.4\,pm and 0.3\,degree for the bond distances and angles, respectively. 
For light main group elements, all methods give a mean absolute deviation of less than 1\,pm as well, which was also observed in previous studies\cite{scand3,caldeweyher2019}.
In comparison with the other methods tested here, r\textsuperscript2SCAN-D4 also yields the lowest MAD of only 0.7\,pm.
Finally, for transition metal complexes r\textsuperscript2SCAN-D4 performs reasonably well with an MAD of 1.9\,pm.
Overall, the performance of r\textsuperscript2SCAN is similar to, and sometimes even better than, the hybrid PBE0-D4, which in turn is one of the best performing hybrid functionals for molecular geometries.\cite{pbeh3c}

\begin{figure*}[htb]
\includegraphics[width=0.48\textwidth]{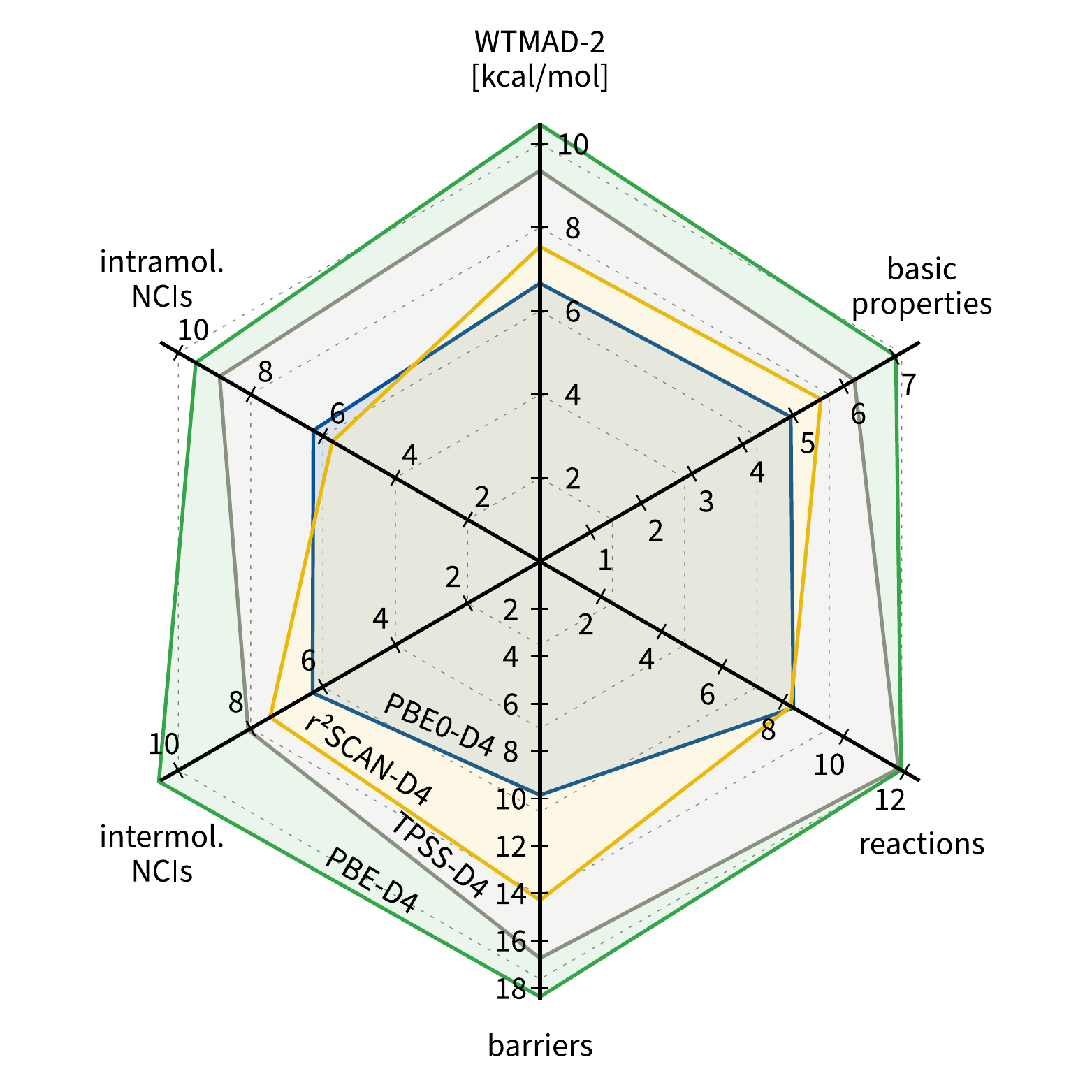}
\includegraphics[width=0.48\textwidth]{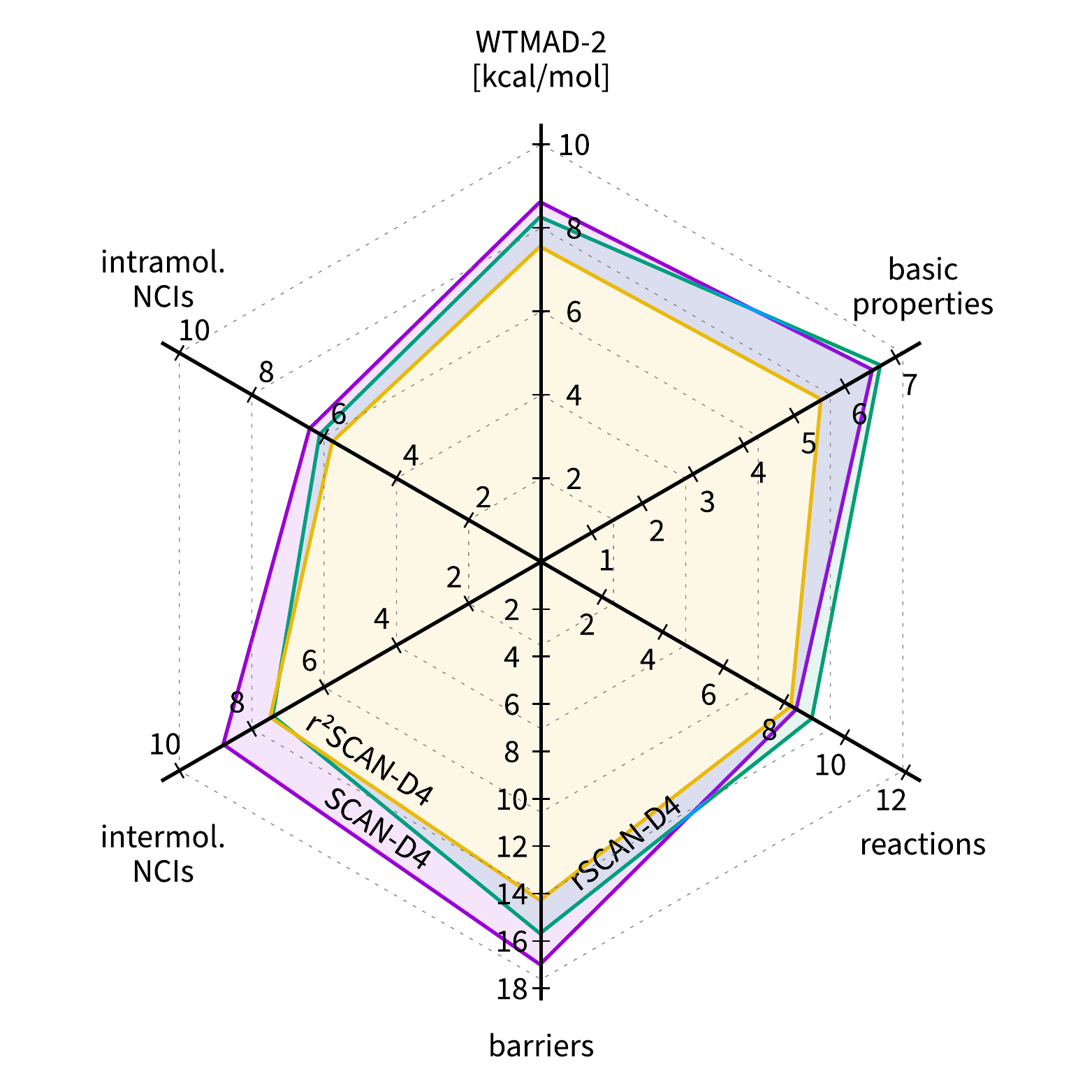}
\caption{\footnotesize Weighted mean absolute deviations of r$^2$SCAN-D4 compared to other DFAs for the large database of general main group thermochemistry, kinetics, and non-covalent interactions GMTKN55~\cite{gmtkn55}.
On the left-hand graphic, r\textsuperscript2SCAN-D4 is compared against functionals representative of their respective rungs.
On the right-hand graphic, r\textsuperscript2SCAN-D4 is compared to other members of the SCAN family, namely rSCAN-D4 and SCAN-D4.
}
\label{fig:gmtkn55}
\end{figure*}

\subsection{\label{subsec:gmtkn55}General main group thermochemistry and non-covalent interactions}

To investigate the performance of r\textsuperscript2SCAN-D4 for general main group chemistry, we use the main group thermochemistry, kinetics and non-covalent interactions (GMTKN55) database.\cite{gmtkn55}
The GMTKN55 database is a compilation of 55 benchmark sets to assess the performance of DFAs and allows a comprehensive comparison of DFAs.
It contains five categories, namely basic properties, barrier heights, isomerisations and reactions, intermolecular, and intramolecular non-covalent interactions~(NCIs).
With the exception of the semi-empirical B97M-V\cite{mardirossian2015} (and its B97M-D4 variant\cite{najibi2020}), r\textsuperscript2SCAN-D4 is the best non-hybrid functional on the GMTKN55 so far with a weighted total MAD (WTMAD2) of 7.5\,kcal/mol.
For the isomerization and reactions category as well as for the intramolecular NCIs, r\textsuperscript2SCAN-D4 can even compete with the performance of the hybrid PBE0-D4.

We additionally evaluated rSCAN-D4 and SCAN-D4\cite{note_scan} on the GMTKN55 set to monitor the development in the SCAN-family of functionals.
The main difference between SCAN-D4 and rSCAN-D4 is the general improvement in the description of non-covalent interactions, while both functionals perform similarly well in all other categories.
Here rSCAN-D4 improves for both NCI categories with a weighted MAD of 6.8\,kcal/mol over SCAN-D4, which yields a weighted MAD of 7.6\,kcal/mol.
This improvement in rSCAN-D4 is mainly responsible for the smaller WTMAD2 of 8.3\,kcal/mol compared to the WTMAD of 8.6\,kcal/mol for SCAN-D4.
For r\textsuperscript2SCAN-D4, the improved description of NCI in rSCAN-D4 is preserved (weighted MAD of 6.6\,kcal/mol) but r\textsuperscript2SCAN-D4 bests its predecessor in all three remaining categories, resulting in its exceptional WTMAD2 of 7.5\,kcal/mol.
The mindless benchmark (MB16-43 subset of GMTKN55) is specifically useful for testing a methods robustness to deal with unusual chemistry in artificial molecules. Here, we see that enforcing exact constraints in non-empirical DFAs yields generally lower errors for artificial molecules than their empirical counterparts (see Table~\ref{tab:mb16-43_results}). 

To stress the importance of including a dispersion correction we test the plain dispersion-uncorrected r\textsuperscript2SCAN which yields a significantly worse WTMAD2 of 8.8\,kcal/mol, a difference similar in magnitude to the improvement from SCAN-D4 to r\textsuperscript2SCAN-D4.
In summary, r\textsuperscript2SCAN-D4 shows a systematic improvement over its predecessor SCAN-D4 in all categories of GMTKN55 and can preserve improvements present in rSCAN-D4.
This makes r\textsuperscript2SCAN-D4 one of the best non-empirical meta-GGAs that have been broadly benchmarked so far.

\begin{table}[bht]
\caption{Comparison of a few non-empirical and empirical dispersion corrected DFAs for the MB16-43 subset (artificial molecules) of GMTKN55. The non-empirical DFAs yield generally lower MADs (in kcal/mol) indicating better transferability across diverse systems.}
\label{tab:mb16-43_results}
\begin{ruledtabular}
\begin{tabular}{lrlr}
Non-empirical DFA & MAD & Empirical DFA & MAD \\ \hline
r\textsuperscript2SCAN-D4 & 14.6 & MN15L & 20.5 \\
SCAN-D4 & 17.3 & M06L &  63.9\\ 
TPSS-D4 & 25.8 & M06L-D4 & 62.6 \\
PBE-D4 & 25.1 & B97M-D4 & 37.5 \\
PBE0-D4 & 16.0 & B3LYP-D4 & 28.4 \\
\end{tabular}
\end{ruledtabular}
\end{table}

\subsection{\label{subsec:mor} Beyond main group chemistry}
Metal organic chemistry is one of the major application areas of non-hybrid DFAs. Here, we use the MOR41 benchmark set that contains 41 closed-shell metal-organic reactions representing common chemical reactions relevant in transition-metal chemistry and catalysis\cite{dohm2018}.
We compare the statistical deviations from high-level references of r\textsuperscript2SCAN-D4 to PBE0-D4, TPSS-D4, and PBE-D4 in Tab.~\ref{tab:mor}.
The r\textsuperscript2SCAN-D4 functional is one of the best meta-GGAs tested so far on the MOR41 benchmark set, with an MAD of 3.3\,kcal/mol.
Compare this to B97M-D4, one of the best meta-GGAs tested on the GMKTN55 set, which yields a larger MAD of 3.8\,kcal/mol\cite{najibi2020}.

\begin{table}[htb!]
    \caption{Reaction energies (kcal/mol) of 41 metal-organic reactions compared to high-level references.\cite{dohm2018}}
    \label{tab:mor}
    
    \centering
    \begin{ruledtabular}
    \begin{tabular}{l*{4}{r}}
         & MD & MAD & SD & AMAX \\
         \hline
         r\textsuperscript2SCAN    &  $2.1$ & $4.4$ & $5.6$ & $17.3$ \\
         r\textsuperscript2SCAN-D4 & $-0.2$ & $3.3$ & $4.3$ & $14.0$ \\
         TPSS-D4                   & $-1.5$ & $3.5$ & $4.4$ & $22.6$ \\
         PBE0-D4                   & $-0.3$ & $2.3$ & $3.1$ & $14.2$ \\
         PBE-D4                    & $-0.1$ & $3.5$ & $4.8$ & $22.7$ \\
    \end{tabular}
    \end{ruledtabular}
\end{table}

\subsection{\label{subsec:nci}Non-covalent interactions in large complexes and molecular crystals}

\begin{figure}[htb!]
\includegraphics[width=0.45\textwidth]{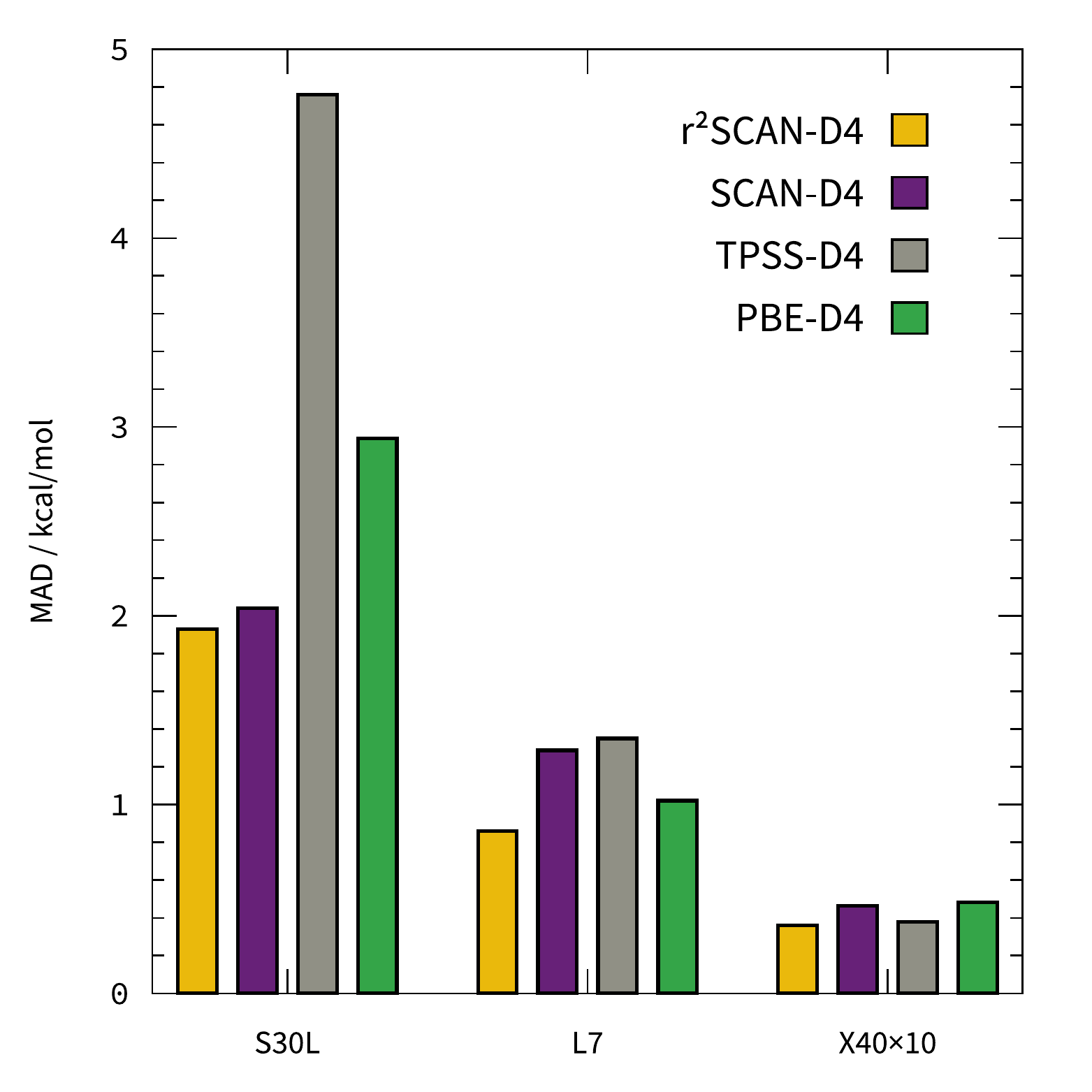}
\caption{\footnotesize Non-covalent interaction energies of host--guest systems, large systems and halogen-bonded systems from r\textsuperscript 2SCAN-D4 compared to high-level references as well as other DFAs.}
\label{fig:nci}
\end{figure}

With the improved description of non-covalent interactions~(NCIs), while retaining the computational efficiency of a meta-GGA, r\textsuperscript2SCAN-D4 is a promising choice for interaction and association energies of large complexes.
The results for the S30L\cite{s30l}, L7\cite{l7} and X40$\times$10\cite{kesharwani2018} benchmark set are shown in Fig.~\ref{fig:nci}.

We choose the recently revised L7 benchmark\cite{l7} set to assess the performance of r\textsuperscript2SCAN-D4 against converged LNO-CCSD(T)/CBS interaction energies\cite{alhamdani2020}.
Close agreement with an MAD of 0.9\,kcal/mol is reached for r\textsuperscript2SCAN-D4.
This is a significant improvement over other meta-GGAs like SCAN-D4 and TPSS-D4 with MADs of 1.3 and 1.4\,kcal/mol, respectively.

We also investigated the description of association energies for large supramolecular complexes using the S30L benchmark set\cite{s30l}.
SCAN-D4 proved to be one of most accurate meta-GGAs in the previous benchmarks \cite{caldeweyher2019}, giving a remarkable MAD of 2.0\,kcal/mol, close to the uncertainty of the provided reference interactions; r\textsuperscript2SCAN-D4 further improves upon this.

In particular, the association energies of the halogen-bonded complexes (15 and 16) are improved with r\textsuperscript2SCAN-D4.
The same trend can be observed in the HAL59 benchmark set of the GMTKN55, which shows an MAD of 1.0\,kcal/mol with SCAN-D4 and improves with r\textsuperscript2SCAN-D4 to an MAD of 0.8\,kcal/mol.
To confirm this trend we additionally evaluated the X40$\times$10 benchmark\cite{kesharwani2018} containing 40 halogen bond dissociation curves with SCAN-D4 and r\textsuperscript2SCAN-D4.
Again, r\textsuperscript2SCAN-D4 gives the lowest MAD of 0.36\,kcal/mol, showcasing on overall improved description of this kind of NCIs.


To evaluate if the good performance for non-covalent interactions transfers from the gas phase to solids, molecular crystals and their polymorphic forms provide useful test cases.\cite{erin_racemic,x23,beran_chemrev}
Here, we investigate the lattice energy benchmark DMC8\cite{dmc8} shown in Table~\ref{tab:dmc8}.
The DMC8 benchmark contains a subset of the X23\cite{c21, x23_ambrosetti,x23_rohl,x23} and ICE10\cite{ice10} benchmark sets with accurate structures and corresponding highly-accurate fixed node diffusion Monte Carlo (FN-DMC) results.
Due to SCAN's tendency to overbind hydrogen bonded systems, like ice polymorphs or hydrogen bonded molecular crystals, dispersion corrected SCAN was problematic for these systems.
With the improved description of non-covalent interactions in r\textsuperscript2SCAN-D4, this issue is mitigated and we find an overall improved MAD of 0.7~kcal/mol.
This MAD is only half of the SCAN-D4 error of 1.5~kcal/mol for these systems and close to the very good performance of the hybrid PBE0-D4 of 0.5~kcal/mol.\cite{caldeweyher2020}
Only the ice polymorphs are systematically overbound by r\textsuperscript2SCAN-D4, which is, however, a problem of many functionals,\cite{ice_ts,wac18} and may be a self-interaction error.\cite{sharkas2020} In contrast, the relative stability of the ice polymorph is reproduced correctly. The energy difference of ice II and ice VIII with respect to ice Ih is 0.03~kcal/mol and 0.70 kcal/mol, respectively, agreeing well with the reference of 0.05~kcal/mol and 0.41 kcal/mol


\begin{table}[htb]
\caption{Lattice energies (kcal/mol) of eight diverse molecular crystals compared to high-level references.~\cite{dmc8} Note the significant improvement from r\textsuperscript2SCAN to r\textsuperscript2SCAN-D4 for the dispersion-bound solids.}
\label{tab:dmc8}
\begin{ruledtabular}
\begin{tabular}{l rrrr}
&ref. & TPSS-D4 & \multicolumn{1}{r}{r$^2$SCAN}& r$^2$SCAN-D4\\[0.1cm]
\cline{2-5}\\[-0.2cm]
Ice\,Ih     & -14.2 & -15.6 & -14.6 & -15.4 \\
Ice\,II     & -14.1 & -14.6 & -14.3 & -15.4 \\
Ice\,VIII   & -13.7 & -12.5 & -13.4 & -14.7 \\
CO$_2$      &  -6.7 &  -5.5 &  -4.7 &  -6.9 \\
Ammonia     &  -8.9 &  -8.6 &  -8.1 &  -9.5 \\
Benzene     & -12.7 & -12.0 &  -5.6 & -12.3 \\
Naphthalene & -18.8 & -18.5 &  -7.5 & -18.6\\
Anthracene  & -25.2 & -24.8 &  -9.9 & -24.7 \\[0.1cm]
\cline{2-5}\\[-0.2cm]
MD          &       &   0.3 &   4.6 &  -0.4 \\
MAD         &       &   0.8 &   5.7 &   0.7 \\
SD          &       &   0.9 &   6.0 &   0.8 \\
AMAX        &       &   1.4 &  15.4 &   1.3 \\
\end{tabular}
\end{ruledtabular}
\end{table}

\begin{figure}[htb]
\includegraphics[width=0.49\textwidth]{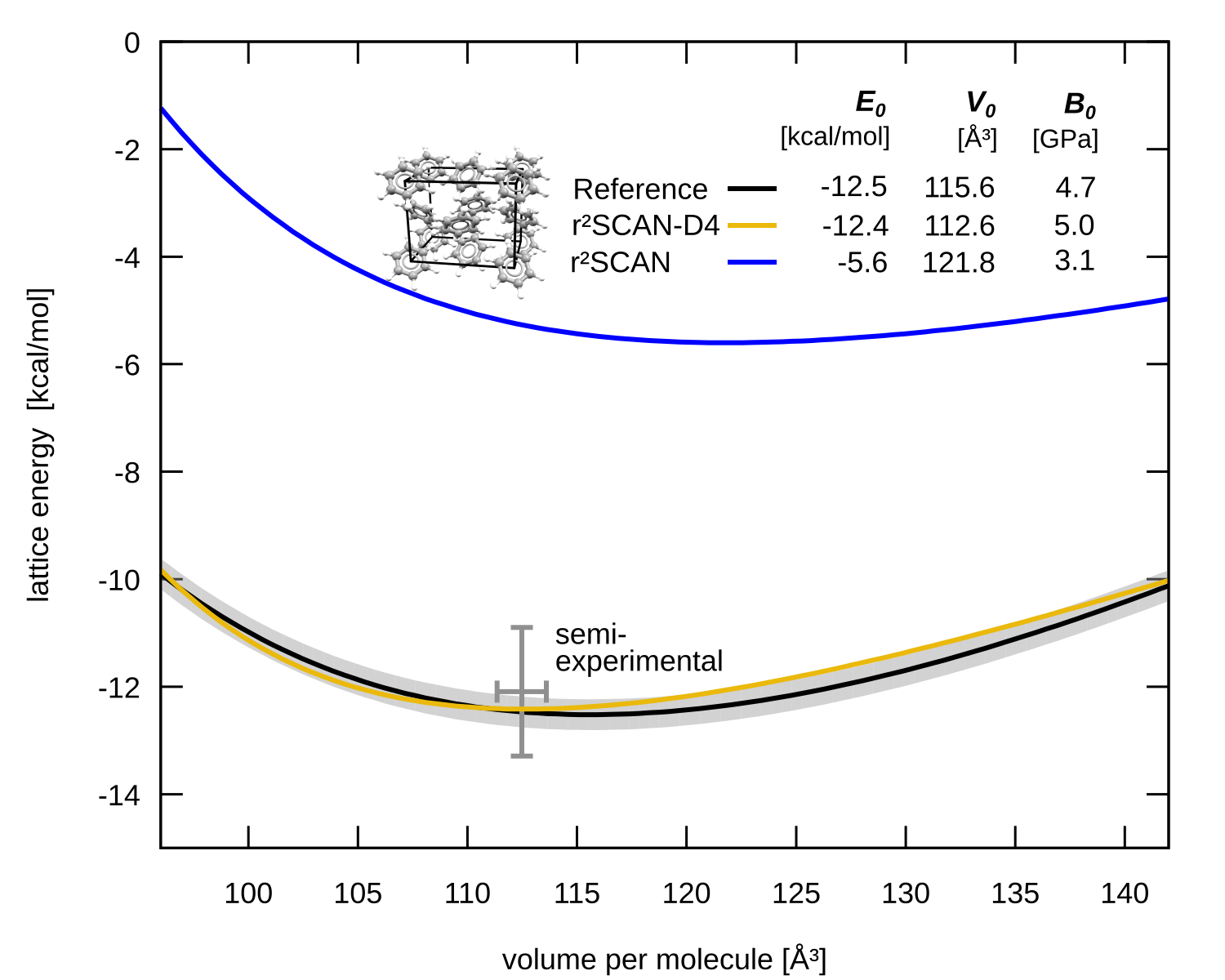}
\vspace{-0.6cm}
\caption{\footnotesize Equation of state for the benzene crystal from r$^2$SCAN-D4 compared to experimental measurements and high-level references taken from Ref.~\citenum{dmc8}.
}
\label{fig:benzene_eos}
\end{figure}

The benzene crystal has been frequently used for electronic benchmark purposes.\cite{Wen:2011gm,sherrill_mbd,631g} Here, we evaluated the equation of state~(EOS) to compare with experimental measurements and the Murnaghan EOS fit to the FN-DMC from Ref.~\citenum{dmc8}.
The resulting EOS is shown in Fig.~\ref{fig:benzene_eos} and agrees excellently with the high-level method as well as the experimental estimate.
A slight underestimation of the unit cell volume by 2.6\% and overestimation of the bulk modulus by 6.4\% can be seen.
To highlight once again the importance of London dispersion on properties beyond the mere energy, we report plain r\textsuperscript2SCAN results as well. The r\textsuperscript2SCAN EOS has a  significant offset equilibrium volume that is overestimated by 5.4\% and a bulk modulus underestimated by 34.0\%.

\section{\label{sec:conclusion}Conclusions}

We have presented an accurate and robust combination of the non-empirical r\textsuperscript2SCAN DFA with the state-of-the-art D4 dispersion correction.
The resulting r\textsuperscript2SCAN-D4 electronic structure method shows exceptional performance across several diverse categories of chemical problems assessed by thousands of high-level data-points in a number of comprehensive benchmark sets.
Included in the assessment were molecular thermochemistry for both main group and transition metal compounds, barrier heights, structure optimizations, lattice energies of molecular crystals, as well as both inter- and intramolecular non-covalent interactions of small to large systems, creating an extensive coverage of chemically relevant problems.

For the large GMTKN55 benchmark collection of about 1500 data points, r\textsuperscript2SCAN-D4 is one of the most accurate meta-GGAs tested so far.
Unlike the best meta-GGA on this set, the dispersion corrected B97M functional, r\textsuperscript2SCAN-D4 can transfer this accuracy to chemically distinct systems like metalorganic reactions.
We find significant improvements in NCIs, which were one of the weak-spots of SCAN based methods. More detailed analysis showed that improvements can mainly be found in the description of hydrogen and halogen bonded systems.
The same trend is found for molecular crystals, where SCAN-D4's tendency to overbind is mostly resolved in r\textsuperscript2SCAN-D4, giving close to hybrid DFT results for lattice energies.

We found r\textsuperscript2SCAN-D4 to be an accurate and (more importantly) consistent DFA for a large variety of problems and chemical systems.
The already good performance of the original SCAN functional is kept and systematically improved in r\textsuperscript2SCAN, while the numeric stability is almost on par with established GGA functionals.
We were able to gain some insight in the improvement from SCAN over rSCAN to r\textsuperscript2SCAN, where we can attribute the improved description of non-covalent interactions to the regularization in the step from SCAN to rSCAN, and the improved thermochemistry and barrier heights to the constraint restoration in the step from rSCAN to r\textsuperscript2SCAN. 
Like SCAN, r\textsuperscript2SCAN is not fitted to molecules, so its accuracy in extensive molecular tests demonstrates the predictive power of its exact constraints and appropriate norms. 

With r\textsuperscript2SCAN-D4, a meta-GGA method is finally available that truly leverages the advantages of its rung in Jacob's ladder, while retaining favorable numerical properties and fulfilling important exact constraints.
We anticipate r\textsuperscript2SCAN-D4 to be a valuable electronic structure method with broad applications in computational chemistry and material science.

\begin{acknowledgments}
\end{acknowledgments}
We thank Stefan Grimme for valuable discussions.
SE is supported by the DFG in the framework of the priority program 1807 ”Control of London dispersion interactions in molecular chemistry”.
JN, and JS acknowledge the support of the U.S. DOE, Office of Science, Basic Energy Sciences Grant No. DE- SC0019350 (core research). ADK acknowledges the support of the U.S. DOE, Office of Science, Basic Energy Sciences, through Grant No. DE- SC0012575 to the Energy Frontier Research Center: Center for Complex Materials from First Principles. JPP was supported by the U.S. National Science Foundation under Grant No. DMR-1939528 (CMMT with a contribution from CTMC).

\bibliography{literature}

\begin{thebibliography}{87}
\bibitem [{not()}]{note_scan}%
  \BibitemOpen
  \href@noop {} {}\bibinfo {note} {For a subset of the GMTKN55 not all of the
  published benchmark results with the SCAN functional were reproducible.
  Therefore, we reevaluated SCAN on the complete GMKTN55.}\BibitemShut {Stop}%
\end{thebibliography}

\end{document}